\documentclass{article}
\usepackage{amssymb}
\begin{document}

\makeatletter
\def\@maketitle{\newpage
 \null
 {\normalsize \tt \begin{flushright} 
  \begin{tabular}[t]{l} \@date  
  \end{tabular}
 \end{flushright}}
 \begin{center} 
 \vskip 2em
 {\LARGE \@title \par} \vskip 1.5em {\large \lineskip .5em
 \begin{tabular}[t]{c}\@author 
 \end{tabular}\par} 
 \end{center}
 \par
 \vskip 1.5em} 
\makeatother
\topmargin=-1cm
\oddsidemargin=1.5cm
\evensidemargin=-.0cm
\textwidth=15.5cm
\textheight=22cm
\setlength{\baselineskip}{16pt}
\title{A New  Noncommutative Product on the Fuzzy Two-Sphere Corresponding to 
         the Unitary Representation of $SU(2)$ and the Seiberg-Witten Map}
\author{  K.~{\sc Hayasaka}\thanks{hayasaka@particle.sci.hokudai.ac.jp}, 
        \ R.~{\sc Nakayama}\thanks{ nakayama@particle.sci.hokudai.ac.jp}
        \  and Y.~{\sc Takaya}\thanks{
                          yuki@particle.sci.hokudai.ac.jp} 
\\[1cm]
{\small
    Division of Physics, Graduate School of Science,} \\
{\small
           Hokkaido University, Sapporo 060-0810, Japan}
}
\date{
  EPHOU-02-006  \\
hep-th/0209240 \\ 
September  2002  
}
%
%
\maketitle

\begin{abstract} 
We obtain a new explicit expression for the noncommutative (star) product 
on the fuzzy two-sphere which yields a unitary representation. 
This is done by constructing a star 
product, $\star_{\lambda}$, for an arbitrary representation of $SU(2)$ which 
depends on a continuous parameter $\lambda$ and searching for the values of 
$\lambda$ which give unitary representations. We will find two series of 
values: $\lambda = \lambda^{(A)}_j=1/(2j)$ and $\lambda=\lambda^{(B)}_j
=-1/(2j+2)$, where $j$ is the spin of the representation of $SU(2)$. 
At $\lambda = \lambda^{(A)}_j$ the new star product 
$\star_{\lambda}$ has poles. To avoid the singularity the functions 
on the sphere must be spherical harmonics of order $\ell \leq 2j$ 
and then $\star_{\lambda}$ reduces to the star product $\star$ obtained 
by Pre\u{s}najder\cite{Pres}. 
The star product at $\lambda=\lambda^{(B)}_j$, to be denoted 
by $\bullet$, is new. In this case the functions on the fuzzy sphere do 
{\em not} need to be spherical harmonics of order $\ell \leq 2j$. The star 
product $\star_{\lambda}$ has no singularity for negative values of $\lambda$ 
and we can move from one representation $\lambda=\lambda^{(B)}_j$ to another 
$\lambda=\lambda^{(B)}_{j'}$ smoothly on the negative $\lambda$ line.  
Because in this case there is
no cutoff on the order of spherical harmonics, the degrees of freedom of 
the gauge fields on the fuzzy sphere coincide with those on the commutative 
sphere.  Therefore, although the field theory on the fuzzy sphere is a system 
with finite degrees of freedom,  we can expect the existence of the 
Seiberg-Witten map between the noncommutative and commutative descriptions 
of the gauge theory on the sphere.  We will derive the first few terms of 
the Seiberg-Witten map for the $U(1)$ gauge theory on the fuzzy sphere 
by using power expansion around the commutative point $\lambda=0$.
\end{abstract}
\newpage
\setlength{\baselineskip}{18pt}

\newcommand {\beq}{\begin{equation}}
\newcommand {\eeq}{\end{equation}}
\newcommand {\beqa}{\begin{eqnarray}}
\newcommand {\eeqa} {\end{eqnarray}}
\newcommand{\bm}[1]{\mbox{\boldmath $#1$}}
\newcommand{\al}{2\pi \alpha'}

\section{Introduction}
\hspace{5mm}
The fuzzy sphere is a 2d surface on which the coordinates are represented by
operators $\hat{x}^a$ $(a=1,2,3)$ that satisfy a condition 
\begin{equation}
(\hat{x}^a)^2 = R^2 {\bm 1}
\end{equation}
as well as a commutation relation
\begin{equation}
\ [\hat{x}^a, \hat{x}^b ] = 2i \lambda \ \epsilon_{abc} \ \hat{x}^c,
\label{algebra}
\end{equation}
where $\lambda$ is a constant parameter, ${\bm 1}$ an identity operator and 
$R$ the radius of the sphere.\cite{Madore}

Because (\ref{algebra}) is the $su(2)$ algebra, for a unitary 
representation $j \ (2j=0,1,2,\ldots)$,  $\hat{x}^a$ is represented 
by a hermitian $(2j+1) \times (2j+1)$ matrix. The $(2j+1)^2$ components of 
this matrix can be decomposed into $2j+1$ irreducible representations of 
$SU(2)$ with angular momentum $\ell=0,1,\ldots,2j$. These representations are 
called spherical harmonics on the fuzzy sphere $\hat{Y}_{\ell m}$ and are
$(2j+1) \times (2j+1)$ matrices.  Therefore the algebra of the spherical
harmonics on the fuzzy sphere is noncommutative. Furthermore in representation 
$j$ the product of $\hat{Y}_{\ell m}$ and $\hat{Y}_{\ell'  m'}$ 
($\ell, \ell' =0,1,\ldots,2j$) 
are expanded in a linear combination of $\hat{Y}_{\ell''  m''}$ with 
$\ell''=0,1,\ldots, \mbox{min}(\ell+\ell',2j)$.   
\begin{equation}
\hat{Y}_{\ell m} \hat{Y}_{\ell' m'} = \sum_{\ell'' m'', (\ell'' \leq 
\mbox{min}(\ell + \ell',2j))} C^{\ell'' m''}_{\ell m,\ell'm'} 
\hat{Y}_{\ell'' m''}
\label{truncation}
\end{equation}
Therefore in contrast to the ordinary spherical 
harmonics $Y_{\ell m}$ there is a truncation of the angular momentum.

Field theories on a fuzzy sphere have been studied by many people.
\cite{Madore}\cite{MSSW}\cite{Watamura}\cite{fuzzyS2}\cite{ARS1}\cite{ain}
In \cite{ARS1} a gauge theory was constructed in the context of
a spherical D2-brane in $SU(2)$ WZW model. The dynamical degrees of freedom are
finite-size matrices.  
In the case of a flat space and infinite matrices the algebra of matrices can 
be realized by that of functions. The multiplication rule of the matrices are 
realized in the algebra of functions by the Moyal product.\cite{DN}
For finite matrices on the fuzzy sphere the corresponding  noncommutative 
(star) product was considered in \cite{MSSW}\cite{Berezin}\cite{BBEW}\cite{Pres}\cite{coherent}\cite{Hoppe}\cite{kishimoto}\cite{HLSJ}.
In \cite{Pres} an explicit formula for the star product on fuzzy sphere  
was constructed using the coherent state method\cite{GroPre}. 
\begin{equation}
f \star g = f \ g + \sum_{m=1}^{2j} \frac{(2j-m)!}{m!(2j)!} 
\ J^{a_1 b_1}J^{a_2 b_2}
\cdots J^{a_m b_m} \ \partial_{a_1} \cdots \partial_{a_m} f \
\partial_{b_1} \cdots \partial_{b_m} g.
\label{starP}
\end{equation}
Here $J^{ab}=x^2\delta^{ab}-x^ax^b+ix\epsilon_{abc}x^c$ and 
$ \ x \equiv \sqrt{(x^a)^2}$. 
The summation stops at $m=2j$ and $f$, $g$ must be polynomials, {\it i.e.}  
spherical harmonics $Y_{\ell m}$ of order $\ell \leq 2j$.
Extension to the fuzzy complex projective space $CP^{N-1}$ was performed 
in \cite{OC}. 
There is also a star product in the integral form. 
In \cite{coherent} by performing the stereographic projection of the sphere on 
to the plane and using generalized coherent states on the complex 
plane\cite{generalizedcoherent} another star product in the integral form 
was constructed. In \cite{iso} this product was also derived from the matrix 
model of \cite{ARS1}. 

In this paper we will derive the following expression for the star
product on the fuzzy sphere.(sec.2)
\begin{equation}
f \star_{\lambda} g = f \ g + \sum_{m=1}^{\infty} C_m(\lambda) 
\ J^{a_1 b_1}J^{a_2 b_2}
\cdots J^{a_m b_m} \ \partial_{a_1} \cdots \partial_{a_m} f \
\partial_{b_1} \cdots \partial_{b_m} g.
\label{starI}
\end{equation}
Here $C_m(\lambda)=\frac{\lambda^m}{m! (1-\lambda)(1-2\lambda) 
\cdots (1-(m-1)\lambda)}$. Note that the summation extends to $m=\infty$.
This product corresponds to an arbitrary representation of $SU(2)$ including 
non-unitary ones. 
$\lambda$ is a parameter introduced in (\ref{algebra}) and this product gives 
the realization of (\ref{algebra}). 
We will show the associativity of this product (\ref{starI}). 
For application to the field theories on the fuzzy two-sphere the values of 
$\lambda$ must be selected  by the condition of unitary representation. 
We will  show  that there exist two values of $\lambda$ for a single unitary 
representation $j$ of $SU(2)$, $\lambda_j^{(A)}=1/(2j)$ and $\lambda_j^{(B)}
=-1/(2j+2)$.(sec.3) For $\lambda=\lambda_j^{(A)}$ the product (\ref{starI}) 
reduces to (\ref{starP}). 
The coefficient $C_m(\lambda^{(A)}_j)=(2j-m)!/m!(2j)!$ 
is singular for $m \geq 2j+1$. To truncate the summation the functions on the 
fuzzy sphere must be polynomials of order $2j$, {\it i.e.}
$f(x)=\sum_{\ell=0}^{2j}\sum_{m=-\ell}^{\ell}a_{\ell m}x^{\ell}Y_{\ell m}$. 
The algebra of the spherical harmonics 
$Y_{\ell m} \ (\ell \leq 2j)$ with the noncommutative product (\ref{starP}) 
coincides with that of the spherical harmonics on the fuzzy sphere 
$\hat{Y}_{\ell m}$. 

The product (\ref{starI}) for $\lambda=\lambda_j^{(B)}$, which we will denote
by $\bullet$, is new and has interesting properties. The coefficient 
$C_m(\lambda^{(B)}_j)=(-1)^m(2j+1)!/m!(2j+1+m)!$ is not singular and there is 
no restriction on the angular momentum of the functions, {\it i.e.} 
$f(x)= \sum_{\ell=0}^{\infty} \sum_{m=-\ell}^{\ell}a_{\ell m}x^{\ell}
Y_{\ell m}$. Furthermore, contrary to the case of the product (\ref{starP}) 
the spherical harmonics $Y_{\ell m}$ with the product $\bullet$ do {\em not} 
realize the algebra of $\hat{Y}_{\ell m}$ explicitly! Especially, a product 
of polynomials of orders $\ell$ and $\ell'$ yields a polynomial of order 
$\ell+\ell'$. It turns out, however, that the integral of the star product of 
$Y_{\ell m}$, $Y_{\ell' m'}^{\ast}$ corresponds to the trace of the product 
of $\hat{Y}_{\ell m}$, $\hat{Y}_{\ell' m'}^{\dagger}$ and the integral 
vanishes for $\ell \neq \ell'$ or $\ell=\ell' > 2j$. 
Therefore just the combination of the star product 
and the integration gives a realization of the fuzzy sphere algebra. 

For noncommutative theories in the flat space, the Seiberg-Witten map 
\cite{SW} between noncommutative and commutative descriptions has been  
investigated.\cite{cohomo}\cite{SWMAP}  
The Seiberg-Witten map is a transformation from the gauge fields in the 
commutative description to those in the noncommutative description 
in such a way that two field configurations in the gauge equivalence class 
in one description are mapped on to the two field configurations in the gauge
equivalence class in the other description. A crucial problem in the study of 
the Seiberg-Witten map for theories on the fuzzy sphere is that while the 
angular momentum of the gauge fields in the noncommutative description is  
at most $\ell=2j$, that in the commutative description is arbitrary. 
Therefore it is difficult to establish a map from the commutative gauge fields 
to the noncommutative gauge fields. Furthermore the order of the polynomials 
will change when we move from $\lambda=\lambda^{(A)}_j$ to $\lambda=0$ along 
the positive $\lambda$ line. However, if we use the star product with 
$\lambda=\lambda^{(B)}_j$, then the order of spherical harmonics  
on the fuzzy sphere are not restricted and we can consider a map 
to commutative gauge fields, although the theory in the noncommutative 
description has only finite degrees of freedom.  
The star product (\ref{starI}) has poles on the positive $\lambda$ line and 
interpolation between $\lambda=\lambda^{(A)}_j$ and $\lambda=0$ along the 
positive $\lambda$ line is not possible.  On the contrary there is no 
singularity on the negative line and the interpolation from $\lambda=0$ to 
$\lambda=\lambda^{(B)}_j$ will make sense. In this sense the second series of 
unitary representations $\lambda_j^{(B)}$ seems  preferable for the study of 
the Seiberg-Witten map.  In sec.4 we will derive the first 
two nontrivial orders ${\cal O}(\lambda^{n}) \ (n=1,2)$ of the Seiberg-Witten 
map by using power expansion around $\lambda=0$. 
We will give a brief summary in sec.5.

\section{Star Product for Arbitrary Representations}
\hspace{5mm}
In \cite{Pres} the star product (\ref{starP}) on the fuzzy two-sphere was 
obtained. This product corresponds to the spin $j$ representation of $SU(2)$. 
The star product (\ref{starI}) can be formally derived by rewriting the 
factorials in (\ref{starP}) by gamma functions and replacing $2j$ inside 
the gamma functions by $1/\lambda$.\footnote{Replacement of $N=2j$ by 
$1/ \epsilon =1/ \lambda$ and power expansion of (\ref{starP}) around 
$\epsilon=0$ was suggested in \cite{Pres}. It was, however, concluded that 
{\em only} the product (\ref{starP}) gives a unitary representation. 
The upper limit of summation was not specified explicitly and there was no 
proof of the associativity for an arbitrary value of $\lambda$.} 
However, additionally, the upper limit of summation, $2j$, must be replaced 
by $\infty$. Therefore 
the associativity of (\ref{starI}) is not {\em a priori} guaranteed. We will 
examine this problem in this section. This product corresponds to arbitrary 
representations of $SU(2)$, including non-unitary ones.  In the next section 
we will select the values of $\lambda$ for unitary representations and find 
a new star product $\bullet$.    

Instead of beginning with (\ref{starI}) we will assume the following form 
for the star product $\star_{\lambda}$. 
\begin{eqnarray}
f(x) \star_{\lambda} g(x)& = & f(x) g(x) + \lambda J^{ab} \partial_a f(x) \ 
\partial_b g(x) \nonumber \\
&&+ \sum_{n=2}^{\infty}\lambda^n \sum_{m=2}^n \chi^{(n)}_{m,m} 
J^{a_1 b_1}J^{a_2 b_2} \cdots J^{a_m b_m} \ \partial_{a_1} \cdots 
\partial_{a_m} 
f \ \partial_{b_1} \cdots \partial_{b_m} g,
\label{productl}
\end{eqnarray}
where $\chi^{(n)}_{m,m}$ $(2 \leq m \leq n)$ is a constant and will be 
determined by the requirement of the associativity. 
Here $\lambda$ was introduced in (\ref{algebra}) and works as an expansion 
parameter. 
We can easily show that this product realizes the algebra (\ref{algebra}).
$J^{ab}$ is a function defined by
\begin{equation}
J^{ab}(x)=x^2\delta_{ab}-x^ax^b+ix\epsilon_{abc}x^c,
\label{J}
\end{equation}
where $\epsilon_{abc}$ is a completely anti-symmetric unit tensor 
($\epsilon_{123}=+1$), and is a projector ($J^{ab}J^{bc}=2x^2J^{ac}$).

First of all we note the following identities of $J^{ab}$.
\begin{enumerate}
\item 
\begin{equation}
x^a \ J^{ab} = x^b \ J^{ab} = 0
\end{equation}
\item 
\begin{equation}
J^{ab} \ J^{ac} = J^{ba} \ J^{ca} = 0
\end{equation}
\item 
\begin{equation}
J^{ab} \ \partial_a J^{cd} = J^{ca} \ \partial_a J^{db}
\end{equation}
\item 
\begin{equation}
J^{ba} \ J^{dc} \ \partial_a \partial_c J^{fe} =
-J^{bf} \ J^{de}-J^{be} \ J^{df}
\end{equation}
\item 
\begin{equation}
J^{a_1b_1}J^{a_2b_2} \cdots J^{a_nb_n} \partial_{b_1} \partial_{b_2}
\cdots \partial_{b_n} \ J^{cd} =0 \qquad ( n \geq 3)
\end{equation}
\end{enumerate}
Identities  1,..5 follow directly from the definition (\ref{J}). 

Identities 1,2 guarantee that $x$ is a constant with respect to the 
$\star_{\lambda}$ product:
{\it i.e.} for any functions $f(x)$ of $x$ and $g(x^a)$ we obtain
\begin{equation}
f(x) \star_{\lambda} g(x^a) = g(x^a) \star_{\lambda} f(x) = f(x) \ g(x^a).
\label{xconstant}
\end{equation}
For example
\begin{equation}
J^{ab}\partial_a f(x) \partial_b g = J^{ab} \frac{x^a}{x}f'(x) \partial_bg=0,
\end{equation}
\begin{equation}
J^{ab}J^{cd}\partial_a\partial_c f(x) \partial_b \partial_d g
= J^{ab}J^{cd} \{ \delta_{ac} f'(x)/x+(x^ax^c/x) \ (f'(x)/x)' \} \partial_b 
\partial_d g =0,
\end{equation}
where $f'(x) \equiv df(x)/dx$.
This is a necessary condition because $\star_{\lambda}$ is the product on the 
sphere. 

Let us check the associativity of (\ref{productl}). To the 1st order 
both $(f \star_{\lambda} g) \star_{\lambda} h$ and 
$f \star_{\lambda} (g \star_{\lambda} h)$ are given by
\begin{equation}
f \ g \ h + \lambda \ J^{ab} \left( \partial_a f \ \partial_b g \ h+ 
\partial_a f \ g \ \partial_b h+ f \ \partial_a g \ \partial_b h\right)
+{\cal O}(\lambda^2)
\end{equation}
and $\star_{\lambda}$ is associative to this order.
To the next order  one can show that the difference 
$(f \star_{\lambda} g) \star_{\lambda} h- f \star_{\lambda} 
(g \star_{\lambda} h)$ is given by 
\begin{eqnarray}
&&\lambda^2 \ (J^{ab} \ \partial_a J^{cd}-J^{ca} \ \partial_a J^{db}) \ \partial_c f \
\partial_d g \ \partial_b h \nonumber \\
&&+\lambda^2 \ (1-2\chi^{(2)}_{2,2}) J^{ac}J^{bd}(\partial_a\partial_b
f \partial_c g \partial_d h-\partial_a f \partial_b g \partial_c \partial_d h).
\end{eqnarray}
The 1st term vanishes due to identity 3 and the other term vanishes if the 
condition 
\begin{equation}
2 \chi^{(2)}_{2,2}-1 =0
\end{equation}
is satisfied. 

To the next order we can show by using identity 4 that the associativity 
imposes the conditions
\begin{equation}
  3 \chi^{(3)}_{3,3} -\chi^{(2)}_{2,2}
= 0, \quad 2\chi^{(3)}_{2,2}-2\chi^{(2)}_{2,2} = 0.
\end{equation}
Therefore we have
\begin{equation}
\chi^{(2)}_{2,2} = \frac{1}{2}, \qquad \chi^{(3)}_{3,3} = \frac{1}{6}, 
\qquad \chi^{(3)}_{2,2}=\frac{1}{2}. 
\end{equation}

We can work out a similar analysis to higher orders. Study of a few more 
higher orders shows that the condition of the associativity leads to the 
recursion relation
\begin{equation}
m \ \chi^{(n)}_{m,m} -m(m-1) \ \chi^{(n-1)}_{m,m}-\chi^{(n-1)}_{m-1,m-1}=0, 
\qquad \chi^{(n)}_{n,n} = \frac{1}{n!}.
\label{recursion}
\end{equation}
Here identities 1-5 must be used. 
The solution to (\ref{recursion}) is not difficult to obtain.
\begin{equation}
\chi^{(n)}_{m,m} = \frac{1}{m!} \sum _{\{P \}}(m-1)^{P_1} \ (m-2)^{P_2} \ 
\cdots 2^{P_{m-2}} \ 1^{P_{m-1}},
\end{equation}
where $P_i$ is the  partition of $n-m$ into $m-1$ nonnegative integers. 
First few examples of $\chi$ can be explicitly written as 
\begin{equation}
\chi^{(n)}_{2,2}=\frac{1}{2}, \quad \chi^{(n)}_{3,3}=\frac{1}{6}(2^{n-2}-1),
\quad \chi^{(n)}_{4,4} = \frac{1}{48}(3^{n-2}-2^{n-1}+1).
\end{equation}
Now the summation over $n$ in (\ref{productl}) can be performed. For
\begin{equation}
C_m(\lambda) \equiv \sum_{n=m}^{\infty} \lambda^n \ \chi^{(n)}_{m,m}
\end{equation}
we immediately obtain 
\begin{equation}
C_2(\lambda) = \frac{\lambda^2}
{2!(1-\lambda)}, \quad C_3(\lambda) = 
\frac{\lambda^3}{3!(1-\lambda)(1-2\lambda)}, \quad 
C_4(\lambda) = \frac{\lambda^4}{4!(1-\lambda)(1-2\lambda)(1-3\lambda)}
\end{equation}
and generally, we get a formula 
\begin{equation}
C_m(\lambda) = 
\frac{\lambda^m}{m! (1-\lambda)(1-2\lambda) \cdots (1-(m-1)\lambda)}.
\end{equation}
We also define $C_1(\lambda)\equiv \lambda$.

In summary we get the associative product $\star_{\lambda}$ on the fuzzy 
two-sphere for arbitrary representation including non-unitary ones.
\begin{equation}
f \star_{\lambda} g = f \ g + \sum_{m=1}^{\infty} C_m(\lambda) \ 
J^{a_1 b_1}J^{a_2 b_2} \cdots J^{a_m b_m} \ \partial_{a_1} \cdots 
\partial_{a_m} f \ \partial_{b_1} \cdots \partial_{b_m} g.
\label{star}
\end{equation}
The coefficient $C_m(\lambda)$ has poles at $\lambda=1,1/2,\ldots,1/(m-1)$
and we must restrict the functional space when we set $\lambda$ to these 
values.  

It is possible to show that with the integration on the sphere this star 
product satisfies the typical 
property of the trace of matrices.
\begin{equation}
\int d\Omega \ f \ \star_{\lambda} g= \int d\Omega \ g \ \star_{\lambda} f
\end{equation}
where $d\Omega$ is the standard measure on the unit sphere. In the polar 
coordinate system it is given by $d\Omega = \sin \theta d\theta d\varphi$. 
Therefore $\int d\Omega$ plays the role of the trace.

\section{Unitary representation of the Fuzzy Sphere Algebra and New Product 
$\bullet$}
\hspace{5mm}
For application to field theory on the fuzzy two-sphere we must pick up 
unitary representations. 
We will determine the allowed values of $\lambda$. 
From the product (\ref{star}) we can derive the fuzzy sphere algebra.
\begin{equation}
 [x^a, x^b]_{\star_{\lambda}} \equiv x^a \star_{\lambda} x^b- x^b 
\star_{\lambda} x^a = 2i \lambda x 
\epsilon_{abc} \ x^c
\end{equation}
The normalized variable $y^a \equiv x^a/(2\lambda x)$ satisfies the standard 
$su(2)$ algebra 
\begin{equation}
 [y^a, y^b]_{\star_{\lambda}} = i \epsilon_{abc} \ y^c
\end{equation}
and for unitary representations Casimir operator $y^a \star y^a$ must take
discrete values $j(j+1)$, $(2j=0,1,2,3, \ldots)$. 
By using (\ref{star}) we get
\begin{equation}
R^2=x^a \star_{\lambda} x^a = (x^a)^2 + \lambda (x^2 \delta_{aa} -(x^a)^2) 
= (1+2\lambda)x^2
\end{equation}
and then we obtain 
\begin{equation}
y^a \star_{\lambda} y^a = (1+2\lambda)/(2\lambda)^2 = j(j+1).
\end{equation}

This equation has two series of solutions. 
\begin{description}
\item [(A)] 
\begin{equation}
\lambda = \lambda_j^{(A)}=\frac{1}{2j}  \qquad (2j=1,2,\ldots),
\end{equation}
\item [(B)]
\begin{equation}
\lambda =\lambda_j^{(B)}= - \frac{1}{2j+2}  \qquad (2j=1,2,\ldots)
\end{equation}
\end{description}

Let us first consider the case (A).  The coefficient is given by 
\begin{equation}
C_m \left(\frac{1}{2j} \right) = \frac{(2j-m)!}{m!(2j)!}
\end{equation}
When $m \geq 2j$, this coefficient is infinite. Therefore the functions 
$f(x)$, $g(x)$ must be polynomials of order at most $2j$. This restriction 
corresponds to the size $(2j+1) \times (2j+1)$ of the matrices in the spin 
$j$ representation. The summation must now be restricted to $0 \leq m \leq 2j$. 
The product (\ref{star}) with this coefficient agrees with the product 
(\ref{starP})
obtained in \cite{Pres}. If we do not restrict the order of $f(x)$, 
$g(x)$, the product (\ref{starP}) does not satisfy the associativity.
The product reduces to the commutative one in the $j \rightarrow 
\infty$ limit, where the matrix size becomes infinite.

Let us turn to the case (B).   The coefficient
\begin{equation}
C_m \left(-\frac{1}{2j+2} \right) = (-1)^m \frac{1}{m!} \ 
\frac{(2j+1)!}{(2j+1+m)!}
\end{equation}
does not diverge for any $m$ and there is no restriction on the functions 
$f$, $g$ on the fuzzy sphere. We will denote the corresponding product by 
$\bullet$.
\begin{equation}
f \bullet g = f \ g + \sum_{m=1}^{\infty}  (-1)^m \frac{1}{m!} \ 
\frac{(2j+1)!}{(2j+1+m)!}\ J^{a_1 b_1}J^{a_2 b_2}
\cdots J^{a_m b_m} \ \partial_{a_1} \cdots \partial_{a_m} f \
\partial_{b_1} \cdots \partial_{b_m} g.
\label{starII}
\end{equation}
This also reduces to the ordinary commutative product for $j \rightarrow 
\infty$.

Let us investigate the case $j=1/2$ in detail.  For the product (\ref{starP}) 
with $\lambda=\lambda^{(A)}_{1/2}$, 
functions must be a linear combination of $1$ and $x^a$ and we get the 
multiplication rule
\begin{eqnarray}
&&x^a \star x^b = x^2 \delta_{ab} + ix \epsilon_{abc}x^c, \nonumber \\
&& 1 \star x^a = x^a \star 1 = x^a.
\end{eqnarray}
This is the algebra of the Pauli matrices $\sigma_a$ and the 2 by 2 unit 
matrix ${\bm I}_2$. $x^a x^b$ does not appear on the RHS of $x^a \star x^b$ and
the angular momentum is truncated.  For the product (\ref{starII}) we get
\begin{eqnarray}
x^a \bullet x^b &=& \frac{4}{3} (x^ax^b-\frac{1}{3}x^2\delta_{ab})
+\frac{1}{9} x^2
\delta_{ab}-\frac{i}{3}x \epsilon_{abc}x^c, \nonumber \\
x^a \bullet x^b \bullet x^c &=& \frac{20}{9} x^a x^b x^c + \cdots 
- \frac{i}{27}x^3 \epsilon_{abc}.
\end{eqnarray}
Therefore the algebra of (Pauli) matrices is not realized.  
When we integrate these products over the sphere, however, we obtain
\begin{equation}
\int d\Omega \ x^a \bullet x^b = \frac{4\pi}{9}x^2 \delta_{ab}, \qquad
\int d\Omega \ x^a \bullet x^b \bullet x^c = -\frac{4\pi i}{27}x^3 
\epsilon_{abc}.
\end{equation}
These correspond to $Tr \sigma_a \sigma_b$, $Tr \sigma_a \sigma_b \sigma_c$.
Therefore the combination  of the star product and the integration realizes 
the matrix multiplication rule. 

We also note that for the $\ell=2$ spherical harmonics, 
$f(x)=\alpha_{ab} x^ax^b$, $g(x)=\beta_{ab}x^ax^b$ 
($\alpha_{ab}$ and $\beta_{ab}$ are symmetric, traceless constant tensors), 
we obtain
\begin{equation}
\int d\Omega (\alpha_{ab} x^ax^b) \star_{\lambda} (\beta_{cd}x^cx^d) 
= \frac{8\pi}{15}\frac{(1+2\lambda)(1+3\lambda)}{1-\lambda}
x^4 \alpha_{ab}\beta_{ab}.
\end{equation}
For $j \geq 1$ and $-1/4 \leq \lambda <0$ this does not vanish.  
However, when $j=1/2$ and $\lambda=-1/3$, this integral vanishes.
As this example shows the space of functions is actually finite dimensional.

In the case (B) the angular momentum is not cut off at some finite value 
and the relation of the algebra of functions to that of $\hat{Y}_{\ell m}$ 
is not clear.
Only after integration the relation of the algebra of functions to that of 
matrices is manifest. 
However, this product {\em does} realize the fuzzy sphere algebra. 
It is known that a star product is not unique. Given a star product $\star$
and a differential operator $\hat{D}$, a new product defined by
\begin{equation}
f \tilde{\star} g \equiv \hat{D}^{-1} \left( (\hat{D}f) \star (\hat{D}g)
\right)
\end{equation}
gives another star product.  Clearly, the relation between $\star$ and 
$\bullet$ is not of this type. We need more understanding of the relation 
between $\star$ and $\bullet$.

\section{Seiberg-Witten Map on the Fuzzy Sphere}
\hspace{5mm}
In this section we will construct the Seiberg-Witten map\cite{SW} for the
 gauge theory on the two-sphere by using perturbation theory in $\lambda$ 
up to order ${\cal O}(\lambda^2)$. 
After expansion we will set $\lambda=\lambda^{(B)}_j$. So we will use 
$\bullet$ for the star product in what follows.  
We will consider only the $U(1)$ gauge theory for simplicity.  

On the fuzzy sphere the gauge 
transformation is defined  \cite{ARS1} by
\begin{equation}
\delta \hat{A}_a=-i L_a \hat{\Lambda} -i[\hat{A}_a,\hat{\Lambda}]_{\bullet}.
\label{gtr}
\end{equation}
Here $L_a$ is the angular momentum operator
$L_a = -i \epsilon_{abc}x^b \partial_c$
and satisfies the relation
$[L_a,L_b]= i \epsilon_{abc}L_c$.
The strength of the gauge field is defined \cite{ARS1} by
\begin{equation}
\hat{F}_{ab}= -iL_a\hat{A}_b+iL_b \hat{A}_a-i[\hat{A}_a,
\hat{A}_b]_{\bullet}-\epsilon_{abc}\hat{A}_c
\end{equation}
and transforms under (\ref{gtr}) by the formula
$\delta \hat{F}_{ab} = i[\hat{F}_{ab}, \hat{\Lambda}]_{\bullet}$.

In commutative gauge theory the gauge field $A_a$ transforms as
$\delta A_a = -i L_a \Lambda$ 
and a field strength
\begin{equation}
F_{ab} = -i L_a A_b + iL_b A_a -\epsilon_{abc}A_c
\end{equation}
is invariant.

We will use the cohomological approach to the Seiberg-Witten 
map.\cite{JMSSW}\cite{cohomo} We introduce the ghost field $C$ and $\hat{C}$.
In the commutative description the BRST transformation is defined by
\begin{eqnarray}
sC=iC^2=0, \nonumber \\
sA_a = -iL_a C, 
\end{eqnarray}
where $s$ is the BRST operator.
Analogously, in the noncommutative description we have
\begin{eqnarray}
s\hat{C} = i\hat{C} \bullet \hat{C}, \nonumber \\
s\hat{A}_a = -i L_a \hat{C} + i [\hat{C}, \hat{A}_a]_{\bullet}.
\label{BRST}
\end{eqnarray}
We will expand $\hat{C}$ and $\hat{A}_a$ in formal power series in $\lambda$ 
whose coefficients are local polynomials in $C$ and $A_a$.
\begin{eqnarray}
\hat{C}= C + \lambda C^{(1)}+\lambda^2 C^{(2)} + \cdots, \nonumber \\
\hat{A}_a = A_a+\lambda A_a^{(2)} + \lambda^2 A_a^{(3)} + \cdots
\label{expansion}
\end{eqnarray}
The fields $\hat{C}$, $\hat{A}_a$ reduce to $C$, $A_a$ at $\lambda=0$.
We will substitute the expansion (\ref{expansion}) into (\ref{BRST}) and 
obtain eqs for $C^{(i)}$, $A^{(i)}$. We get
\begin{eqnarray}
sC^{(1)}&& = -i \epsilon^{abc}xx^c\partial_a C \partial_b C, \qquad 
sA^{(2)}_a = -iL_a C^{(1)} -2\epsilon_{bcd}xx^d\partial_b C \partial_c A_a,
\nonumber \\
 sC^{(2)}&&= -2\epsilon_{abc}xx^c \partial_a C\partial_b C^{(1)}  
-\epsilon_{abc}xx^c(x^2\delta_{de}-x^dx^e)\partial_a \partial_d C \partial_b 
\partial_e C, \nonumber \\
 sA_a^{(3)}&&=-iL_a C^{(2)}-2\epsilon_{bcd}xx^d (\partial_b C 
\partial_c A^{(2)}_a + \partial_b C^{(1)} \partial_c A_a) \nonumber \\
&&+2x(x^2\delta_{bc}-x^b x^c)\epsilon_{def}x^f 
\partial_{b}\partial_{d}A_a\partial_{c}\partial_{e}C.
\label{1storder}
\end{eqnarray}

The solution to the 1st eq of (\ref{1storder}) is given by 
\begin{equation}
C^{(1)} = \frac{i}{x} \epsilon^{abc} x^a A_b L_cC.
\label{C1}
\end{equation}
Similarly $A^{(2)}_a$, $C^{(2)}$, $A^{(3)}_a$ can be obtained.  
Only the results are presented here. 
\begin{eqnarray}
C^{(2)} &=& - \frac{i}{x^2} \epsilon_{abc} x^b x^d A_cA_dL_aC 
- \frac{i}{x^2} \epsilon_{abc} \epsilon_{def}x^bx^{d}A_c
F_{ae}L_fC \nonumber \\ &&
-\frac{1}{x^2}\epsilon_{abc}\epsilon_{def} x^bx^{d}A_cA_e L_f L_a C
+\frac{i}{x}L_a A_d \epsilon_{abc} x^b L_{c}L_{d}C \ 
+\frac{i}{x}A_a \epsilon_{abc}x^bL_cC, \nonumber \\
A^{(2)}_a &=& \frac{1}{x} \epsilon_{bcd}x^c A_{d}(iL_bA_a+F_{ab}), \nonumber \\
A^{(3)}_a &=&
L_a \left(
\frac1{x^2}x^c A_c (x^d A_d L_b A_b +i A_b F_{bd}x^d)
+\frac1{x}x^c A_c L_b A_b 
-\frac1{x}\epsilon_{bcd}F_{be}x^cL_dA_e
\right) \nonumber \\
&&-3 F_{ab}F_{bc}A_c 
+A_c A_d L_c L_d A_a 
-A_c A_c(L_bL_b A_a -i L_b F_{ab}) \nonumber \\ &&
-2i F_{ab}A_b L_c A_c
+2i F_{ab} A_c L_c A_b
-i A_b L_b F_{ac} A_c -i L_b A_a F_{bc} A_c \nonumber 
\\ &&
+\frac{i}{x}\epsilon_{bcd}x^b L_c L_e A_a L_e A_d
-\frac1{x}L_e F_{ab} \epsilon_{bcd} x^c L_d A_e 
\label{A3}
\end{eqnarray}
In this way one can compute the Seiberg-Witten map for the gauge theory on 
the two-sphere order by order in perturbation theory in powers of 
$\lambda \ ( \ =\lambda^{(B)}_j)$.

\section{Summary}
\hspace{5mm}
In this paper the star product $\star_{\lambda}$ for the field theories 
on the fuzzy sphere corresponding to an  arbitrary representation of 
$SU(2)$ including non-unitary representation is constructed and the 
associativity of this product is proved. By imposing the condition of unitary 
representation we obtained a new noncommutative product $\bullet$ for 
$\lambda=\lambda^{(B)}_j$. 
This new product has a novel feature that the product of spherical harmonics 
with the product $\bullet$ does {\em not} realize the multiplication rule of 
the 
matrices, {\it i.e.} spherical harmonics on the fuzzy sphere 
but after integration over the sphere the result agrees with the trace of the 
product of the corresponding matrices.  The functions on the sphere are not 
restricted to be spherical harmonics of order $2j$ ($j$ is the spin of the 
representation) but the integration eliminates spherical harmonics of order  
larger than $2j$.

The product $\star_{\lambda}$ has singularities on the positive $\lambda$ 
line, while there is no singularity for negative $\lambda$.  Therefore if 
we want to move from one unitary representation $j$ to another $j'$ passing 
through non-unitary representations, especially if we consider the 
Seiberg-Witten map, the negative value $\lambda=\lambda^{(B)}_j$ seems 
suitable.  
Along this move the gauge fields can be spherical harmonics of any order. 
We derived the first few terms of the Seiberg-Witten map by power expansions 
around $\lambda=0$. To derive an exact form of the map without the power 
expansions like that derived in the flat case \cite{SWMAP} a further 
investigation is necessary. 
An investigation such as that of the correspondence 
of the Dirac-Born-Infeld actions in the noncommutative and commutative 
descriptions will be reported elsewhere.

\section*{Acknowledgments}
\hspace{5mm}
The work of R.~N. is supported in part by Grant-in-Aid (No.13135201) 
from the Ministry of Education, Science, Sports and Culture of Japan 
(Priority Area of Research (B)(2)).

\section*{Note Added}
\hspace{5mm}
\begin{description}
\item [(1)] We can prove the following  formula for the inner product of 
ordinary spherical harmonics.  
\begin{equation}
\int d\Omega \ (Y_{\ell m})^* \star_{\lambda} Y_{\ell' m'} = 
\prod_{k=1}^{\ell} \left( \frac{1+(k+1)\lambda}{1-(k-1)\lambda} 
\right) \ \delta_{\ell \ell'} \delta_{mm'}
\label{innerprod}
\end{equation} 
Especially, for $\lambda=\lambda^{(B)}_j$ we obtain
\begin{equation}
\int d\Omega \ (Y_{\ell m})^* \bullet Y_{\ell' m'} = 
\prod_{k=1}^{\ell} \left(\frac{2j-k+1}{2j+k+1}\right) \ \delta_{\ell \ell'} 
\delta_{mm'}. 
\end{equation} 
The inner product is positive definite for $\ell =0,1, \ldots, 2j$ and 
the norm vanishes for $\ell > 2j$.  This shows that our new star product  
$\bullet$ with $\lambda=\lambda^{(B)}_j$  defines a unitary representation.  
This also completes the observation made around eq (41) concerning the finite
dimensionality of the functional space. 

The proof of eq(\ref{innerprod}) will be presented elsewhere.  

\item [(2)] Generally, the solution to  (\ref{1storder}) is not unique and 
has ambiguity of $s$-exact
terms.  We also found a simpler solution.
\begin{eqnarray}
C^{(2)} &=& -2 \epsilon_{abc}\epsilon_{def}\frac{x^cx^f}{x^2} \ 
A_a L_b A_d L_e C \nonumber \\ &&
-2i \epsilon_{abc}\frac{x^cx^d}{x^2} \ A_a A_d L_b C 
- i \epsilon_{abc}\frac{x^c}{x} L_d A_a L_d L_b C, \nonumber \\
A^{(3)}_a &=& 2\epsilon_{bcd} \epsilon_{efg} \frac{x^dx^g}{x^2} A_b 
 \left( -L_c A_e L_f A_a + F_{ce} F_{fa} -i A_e L_c F_{fa}
\right) \nonumber \\
&&+2 \epsilon_{bcd} \frac{x^d x^e}{x^2} A_b A_e (F_{ca} -iL_c A_a) 
+\epsilon_{bcd}\frac{x^d}{x} L_e A_b L_e(F_{ca}-iL_c A_a)
\end{eqnarray}
$C^{(1)}$ and $A^{(2)}_a$ are the same as in (\ref{C1}), (\ref{A3}).

\end{description}



\end{document}